# BEAM LOSS MONITORING WITH FIXED AND TRANSLATING SCINTILLATION DETECTORS ALONG THE FERMILAB DRIFT-TUBE LINAC*

J. Stanton, R. Sharankova, K. Seiya, M. Wesley
Fermi National Accelerator Laboratory, 60510 Batavia, USA

*Abstract*

The Fermilab Linac is a roughly 145 meter linear accelerator that accelerates H- beam from 750 keV to 400 MeV and provides beam for the Booster and the rest of the accelerator chain. The first section of the Linac is a Drift-Tube Linac (DTL), which in its current state, suffers from a lack of instrumentation along its length. As a result, operational staff do not have access to the diagnostic information needed to tune the critical components of this accelerator, such as the quadrupole magnets within the drift tubes. This work presents an effort to utilize both fixed and translating scintillation detectors to investigate beam loss along the first two cavities of the DTL.

## INTRODUCTION

The Fermilab DTL consists of five cavities, referred to as tanks, that span a length of roughly 80 meters and accelerate beam from 750 keV to 116 MeV. There are 207 drift tubes spread throughout the DTL and each one contains a quadrupole magnet inside. A survey of the drift tube positions that was conducted from 2018 to 2022 found that the vertical positions of the drift tubes have a maximum offset of 3.5 mm and the horizontal positions have a maximum offset of 1.8 mm from the tank centers [1]. Tuning of these magnets is not common practice due to a lack of instrumentation along the length of the DTL. This work demonstrates the use of scintillation detectors to measure beam loss along the first two tanks of the DTL with the ultimate goal of using them as a diagnostic tool for tuning the quadrupole magnets.

## DETECTOR INSTALLATION AND SIGNAL

The detectors used in this study consisted of the following components: a photomultiplier tube, a light guide, and a plastic scintillator plate. A large scintillator plate (roughly 1000 cm$^2$) detector, or paddle, was used for translating measurements, while several small scintillator plate (less than 60 cm$^2$) detectors were used for fixed position measurements.

### Translating Cart Design

A simple translating cart, track and pulley system was designed to hold the paddle and allow for it to be pulled back and forth along a 2.3 meter length of the west side of Tank 1. The track was positioned such that the paddle's scintillator plate would remain roughly 40 cm away from the outside of the tank while translating. Figure 1 displays the CAD model of the paddle mounted in the translating cart on the track as well as the pulley system which doubled as buffer stop to prevent the cart from rolling off of the track.

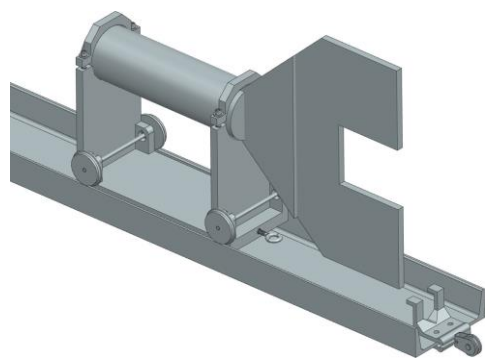

Figure 1: Paddle detector on the translating cart and track with the pulley and buffer stop assembly.

### Detector Configuration and Signal

All detectors were positioned such that their scintillator plates were roughly aligned with the beam orbit. The signal from the detectors, and all other diagnostics was viewed and recorded using an oscilloscope. The oscilloscope was triggered off of a beam timing signal. The raw waveform from the detectors exhibited a substantial amount of noise, from the radio frequency waves (RF) in the tanks, which obscured the signal of interest. This noise was mitigated by implementing a choke on the signal cable, a 500 kHz low pass filter on the oscilloscope input and averaging every 16 samples.

The detector signal structure consisted of a rising and falling edge connected by a flat top, which is referred to as a "shoulder". A step up to a higher flat top when beam was present was also observed. To better understand the cause of the shoulder structure, measurements of the detector signal, the Tank 1 RF gradient and the Tank 1 entry toroid were taken simultaneously in two scenarios. The first scenario injected beam asynchronously with respect to when the RF is pulsed, and the other, injected beam synchronously with the RF pulse as shown in Figure 2.

In both measurements, the shoulder in the detector signal was synchronous with the RF gradient signal. This data indicates the shoulder is caused by the detector picking up X-rays generated by the RF in the tank and is not related to beam.

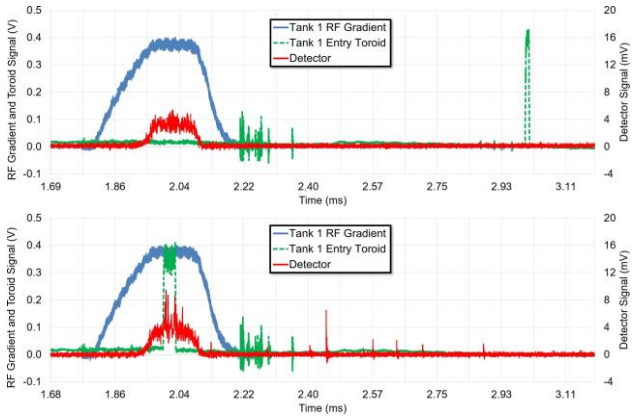

Figure 2: Plots of the detector, Tank 1 RF gradient and Tank 1 entry toroid signals in two scenarious, RF pulse asynchronous with beam (top plot) and RF pulse synchronous with beam (bottom plot).

The additional step from the shoulder to the maximum of the signal, which took place when beam is present, illustrated that the detectors are capable of detecting the products of beam loss.

## TRANSLATING DETECTOR RESULTS

The paddle detector was pulled, in 10 cm increments, along a 2.3 m length of Tank 1. At each measurement location, an upstream horizontal dipole corrector (L:MDQ2H) was varied between three current settings to tilt beam trajectory and induce scraping loss. For each measurement, waveforms of the paddle detector and Tank 1 entry toroid were recorded at a sample rate of 5 mega samples per second. The toroid signal was used to help identify the time in which beam was present in the tank. Using this information, the paddle detector signal could be broken up into two different time spans called the beam and shoulder windows as shown in Figure 3. The length of these windows and their timing with respect to the trigger of the oscilloscope was held constant.

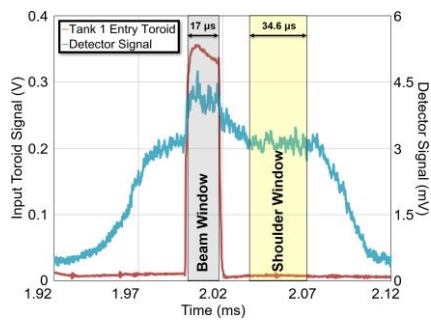

Figure 3: Plot of a single measurement with the beam window and shoulder window highlighted.

Averages of the detector signal over the beam and shoulder windows were calculated for three dipole corrector current settings at 24 measurement positions. Figure 4 is a plot of these averages. Error bars represent plus or minus one standard deviation in the detector signal over each window.

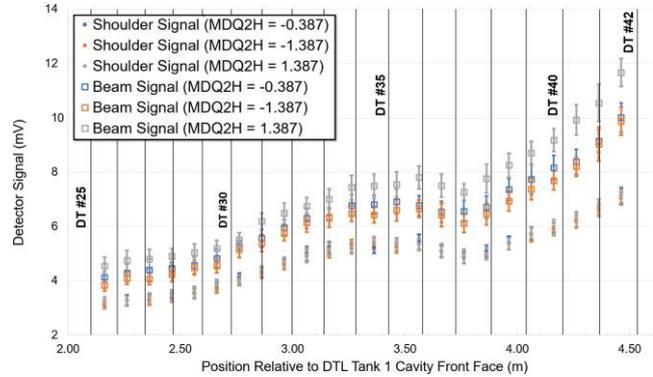

Figure 4: Plot of the beam window and shoulder window averages, for three different dipole corrector settings at 24 longitudinal positions along Tank 1. Drift tube positions are the black vertical lines.

Figure 4 shows that both the beam and shoulder window averages scale with longitudinal position with the exception of the dip in all traces that takes place near 3.5 m. Upon further inspection, this was found to be correlated with when the scintillator plate of the paddle was adjacent to a steel alignment fixture on the side of the tank. This additional material would have blocked any products of the RF and beam induced loss leading to the observed dip. An interesting characteristic of this plot is that the beam window averages scale with dipole corrector current while the shoulder window averages do not. This indicates that the shoulder signal is not related to beam. Furthermore, Tank 1 was designed such that the average field gradient is not constant along its length. Over the length in which the paddle was pulled, the average field gradient increases from 5.51 MV/m to 5.82 MV/m as shown in Figure 5. As the average field gradient increases, so does the contribution of the RF induced X-rays to the detector signal.

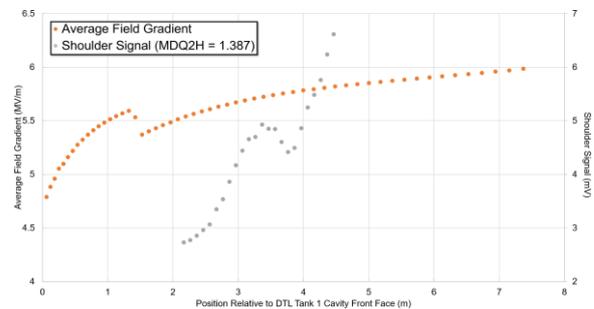

Figure 5: Plot of average shoulder signal with dipole corrector setting equal to 1.387 A and the average field gradient in Tank 1.

The difference between the beam window and shoulder window averages in Figure 4 is not covered by the error in either. This shows that the paddle detector is capable of detecting the products of beam loss. To decouple the effects of RF and beam on the paddle detector signal, the shoulder window averages were subtracted from the beam window averages as an offset. The result, called the paddle signal delta, is plotted in Figure 6. Error in the paddle signal delta is propagated through as the square root of the sum of the squares of the standard deviations in the beam and shoulder window data.

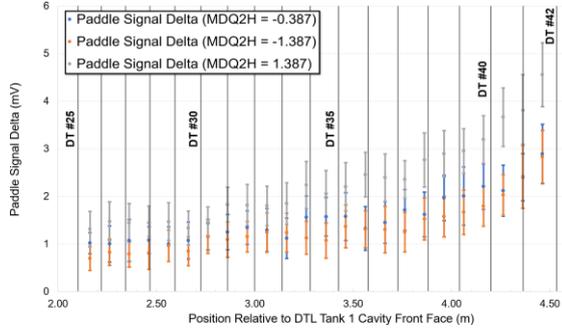

Figure 6: Plot of a the paddle signal delta for three different dipole corrector settings at 24 longitudinal positions along Tank 1. Drift tube positions are the black vertical lines.

The beam toroid signals were also measured to estimate beam loss. This was done by calculating the difference between the average beam current measured at the Tank 1 entry toroid (L:TO1IN) and the Tank 3 entry toroid (L:TO3IN) at each dipole corrector current setting, as shown in Table 1.

Table 1: Beam Loss During Translating Detector Scan

| Parameter | Value | Error |
| --- | --- | --- |
| L:TO1IN Average | 19.85 mA | 0.13 mA |
| L:TO3IN Average | 11.52 mA | 0.14 mA |
| Beam Loss Average | 8.33 mA | 0.19 mA |
| Transmission Efficiency | 58.03 % | 0.81 % |

(a) L:MDQ2H = 1.387 A

| Parameter | Value | Error |
| --- | --- | --- |
| L:TO1IN Average | 22.09 mA | 0.11 mA |
| L:TO3IN Average | 20.76 mA | 0.11 mA |
| Beam Loss Average | 1.33 mA | 0.16 mA |
| Transmission Efficiency | 93.96 % | 0.69 % |

(b) L:MDQ2H = -0.387 A

| Parameter | Value | Error |
| --- | --- | --- |
| L:TO1IN Average | 21.16 mA | 0.09 mA |
| L:TO3IN Average | 19.58 mA | 0.1 mA |
| Beam Loss Average | 1.57 mA | 0.14 mA |
| Transmission Efficiency | 92.57 % | 0.63 % |

(c) L:MDQ2H = -1.387 A

From Table 1a it is clear that the highest beam loss of 8.33 mA was recorded when the dipole corrector current setting was equal to 1.387 A. The paddle signal delta values for this dipole corrector setting are represented by the gray trace in Figure 6. The paddle signal delta values for this trace are higher than the values for the other two dipole corrector settings at every longitudinal position which is expected due to the higher measured beam loss in Table 1a. However, the errors in these values overlap at most positions which makes this result inconclusive in most cases. The separation between the gray trace and the other two increases with position. Lengthening the translation range of the paddle may reveal a more significant difference in paddle signal delta values at higher beam energies, further down the tank.

## STATIONARY DETECTOR RESULTS

Data was also taken from two small scintillator plate detectors that were positioned on either side of the downstream end of Tank 2. The signals captured were sloped in opposite directions, as shown in Figure 7. This indicates that the beam pulse is tilting from head to tail horizontally, resulting in an asymmetrical loss pattern and, as a result, asymmetry in the detector signals.

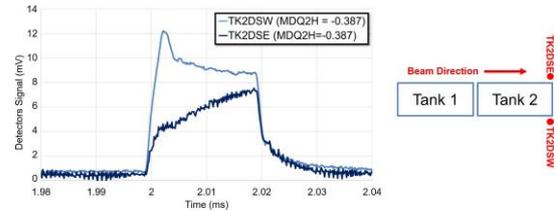

Figure 7: Plot of both stationary detector signals, exhibiting opposite slopes.

## CONCLUSION

This work highlights an effort to utilize both fixed and translating scintillation detectors to investigate beam loss along the first two tanks of the Fermilab DTL. The detectors used in this study were found to be capable of detecting the products of beam loss. However, it is still not conclusive if the paddle delta signal, which is correlated with beam loss, can be measured accurately enough to distinguish between small changes in upstream equipment for the purpose of machine tuning. Current efforts will be focused on lengthening the paddle detector's translation range and it's positional accuracy, completing simulation work to determine what the loss products being detected are and also digitizing the detector signals for integration with Fermilab's accelerator control system.